\documentclass[preprint,12pt]{elsarticle}
\usepackage{lineno}
\linenumbers

\usepackage{graphicx,hyperref}
\usepackage{wrapfig}
\usepackage{epsfig}

\usepackage{amssymb}

\begin{document}

\begin{frontmatter}



\title{SalSA Status}


\author[label1]{Amy Connolly\fnref{label2}}
\ead{connolly@mps.ohio-state.edu}
\author{ \emph{for the SalSA Collaboration}}
\address[label1]{1040 Physics Research Building, 191 West Woodruff Avenue, Columbus, Ohio 43210-1117  USA}
\fntext[label2]{Previous to October, 2010, at the University College London.}

\begin{abstract}
We report on the status of the Salt Sensor Array (SalSA), a proposed
experiment for detecting ultra-high energy neutrinos through 
the radio \v{C}erenkov technique with an array of radio-microwave 
antennas embedded in a large, naturally occurring salt formation.
We review the measurements to date aimed
at assessing SalSA's feasibility, including 
a return visit of the Hockley Salt Mine in Hockley, Texas, and
discuss the current status of the project.
\end{abstract}


\end{frontmatter}


\section{Introduction}
\label{sec:intro}
The Salt Sensor Array (SalSA) is a proposed antenna array
 aiming 
 to detect the well sought after 
ultra-high energy (UHE) ``BZ'' neutrino flux
via the radio \v{C}erenkov signal induced by neutrino-induced
particle cascades in the salt~\cite{hockley,saltaccelerator}.  
It would be deployed in one of the many naturally
occurring 10's-of-km$^3$-scale 
salt formations called diapirs that 
exist throughout the world.
Salt diapirs are formed from $\sim$10~km deep salt beds 
originating from 100-200~million year old dried sea salt.
The salt is buoyed upward over geological time scales 
due to pressures induced by the density
gradient between salt and the surrounding rock.  
Salt domes typically have purities of approximately  
95\%~\cite{brianlock} and pure salt is expected to have long attenuation
lengths in the radio/microwave frequency range.  In addition,
with the ratio of densities being 2.4/0.9, one finds a proportionately
larger number nuclear targets in a similar volume of salt.
These points lead to the
idea that a sparse antenna array could be deployed in salt 
for a cost-effective method of detecting the rare ultra-high
energy neutrino flux.  SalSA could be competitive if 
an array
of order 100 antennas could be deployed in salt 
with attenuation
lengths greater than about 250 meters at 200~MHz.
\vspace*{-0.25in}

Since its inception SalSA has been in the feasibility stage,
awaiting a confirmation that the attenuation lengths in
naturally occurring salt are indeed long enough and sensor
deployment practical and inexpensive 
enough for SalSA to be competitive.
Drilling into a salt dome from the surface 
through the cap rock, however, has been found to be prohibitively expensive,
with costs on the order of \$1M/hole driven by the price of 
drilling for oil with the same machinery.  Nevertheless there are
alternative scenarios, for example where deep holes
are drilled into the salt from within an existing mine.

Here we review the measurements of radio
attenuation in salt {\it in situ} that have
been made to date, with a focus on recent measurements in
the Hockley Salt Mine.  We 
conclude with remarks on the status of
the project.

\section{Radio Attenuation Lengths in Salt}

\label{sec:measurements}

Ground penetrating radar (GPR) measurements have long pointed to 
long attenuation lengths in naturally occurring salt.
Since GPR measurements are focused on timing rather than on power loss,
it is difficult to deduce exact attenuation lengths from their results,
but 
a 1976 paper by Stewart and Unterberger~\cite{s-unterberger} 
reports signals detected
after a 4080~ft. round trip journey through the Cote Blanche Salt Mine in
New Iberia, Louisiana.  In~\cite{ourpaper}, Connolly {\it et al.} 
estimate an implied lower limit on the field attenuation length $L=140$~m
(194~m ice equivalent) at frequency $f=440$~MHz in Cote Blanche salt.  For
a constant loss tangent, $L\propto 1/f$~\cite{hockley}.

The first dedicated {\it in situ} attenuation length measurements in 
salt were made at the Hockley Mine in Hockley, Texas  
using a 10-100~ns pulse, modulated 
with frequencies from 90-500 MHz.  Signals were received over
distances as long as 40~m with 
antennas resonant at 150, 300 and 750~MHz
placed against the walls 
in the mine at 
460~m depth~\cite{hockley}.  
Their results are consistent with 40-300~m
attenuation
lengths in the Hockley mine.
The signal strengths seen at the furthest distances 
indicated little or no attenuation.

Motivated by the promising GPR results,
a team led by Connolly followed with {\it in situ} measurements at Cote Blanche~\cite{ourpaper}.
They used a 2.5~kV pulser with a 10\%-90\% rise time of 200~ps 
transmitted from many different 
dipole antennas spanning 50-900~MHz.   They transmitted and received the signals
within boreholes
at 100-200~ft. depths below the 1500~ft. deep level of the mine,
over distances as great as 169~m.  Their 2009 result 
is the most precise measurement of its kind to date and shows
attenuation lengths of $93\pm7$~m (129~m ice equivalent) at 150~MHz
and $L\propto f^{-0.57\pm0.04}$.

In 2009, armed with the experience from the Cote Blanche visits
and the same high voltage system,
a team consisting of many of the same people arranged to make 
a return visit
to the Hockley mine to reduce the uncertainties on the Hockley
result.  The team included Dr. A. Connolly, Dr. S. Bevan,
Matthew Mottram and Dr. R. Nichol from University 
College London, A. Goodhue and Prof. D. Saltzberg from
UCLA, and 
C. Miki and Prof. P. Gorham from the University of Hawaii, Manoa.  
Bevan, Mottram, 
Nichol and Goodhue performed the measurements in the mine. 
Again they used dipole antennas spanning the frequency
range between 50 and 700~MHz.

\begin{figure*}[h]
\begin{minipage}{15pc}
 \centering
   \includegraphics[width=15pc]{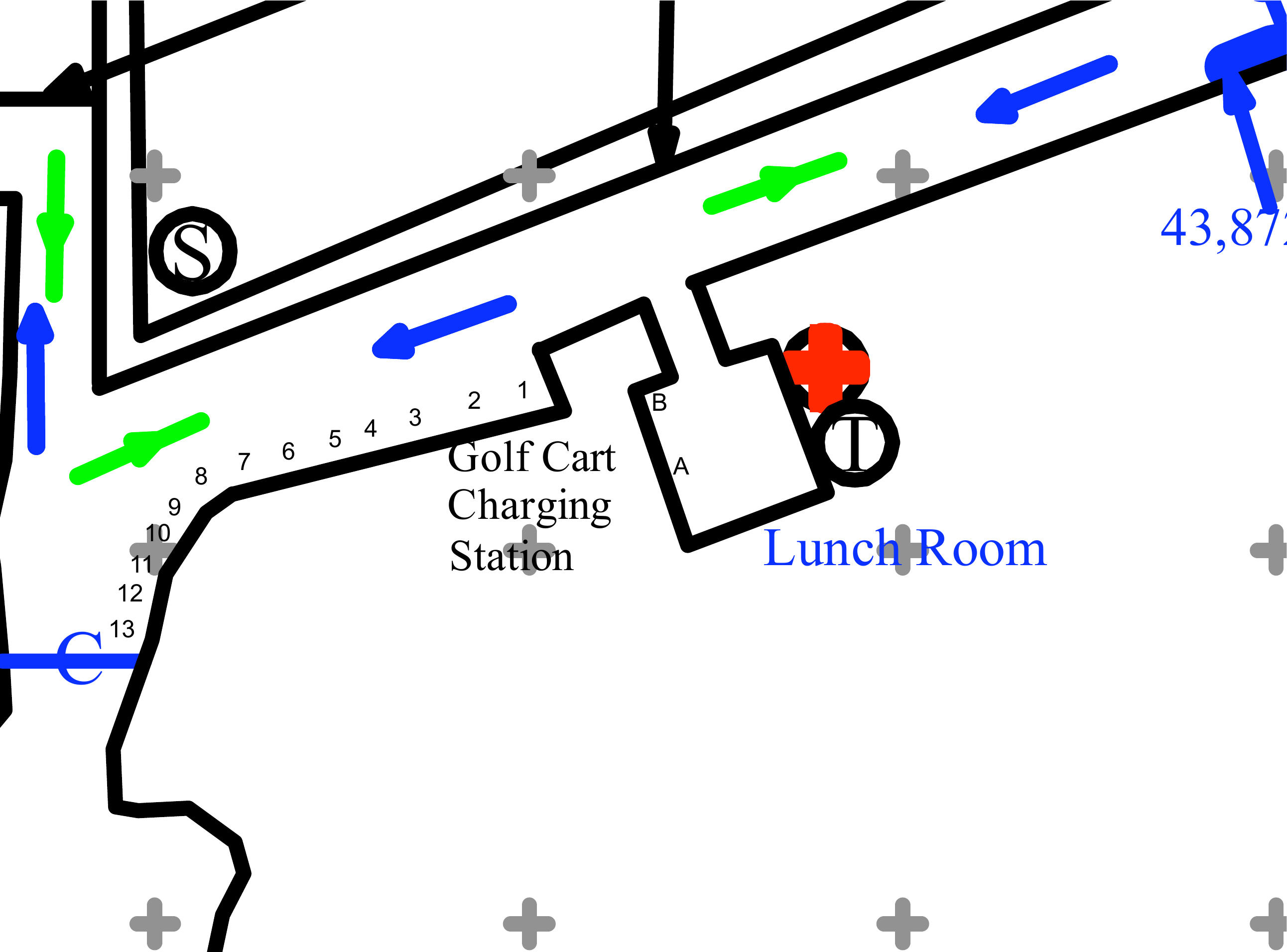}
    \caption{Map of area in Hockley mine 
where measurements were made in 2009.
      \label{plot:1}}
\end{minipage}\hspace{2pc}%
\begin{minipage}{15pc}
 \centering
   \includegraphics[width=15pc]{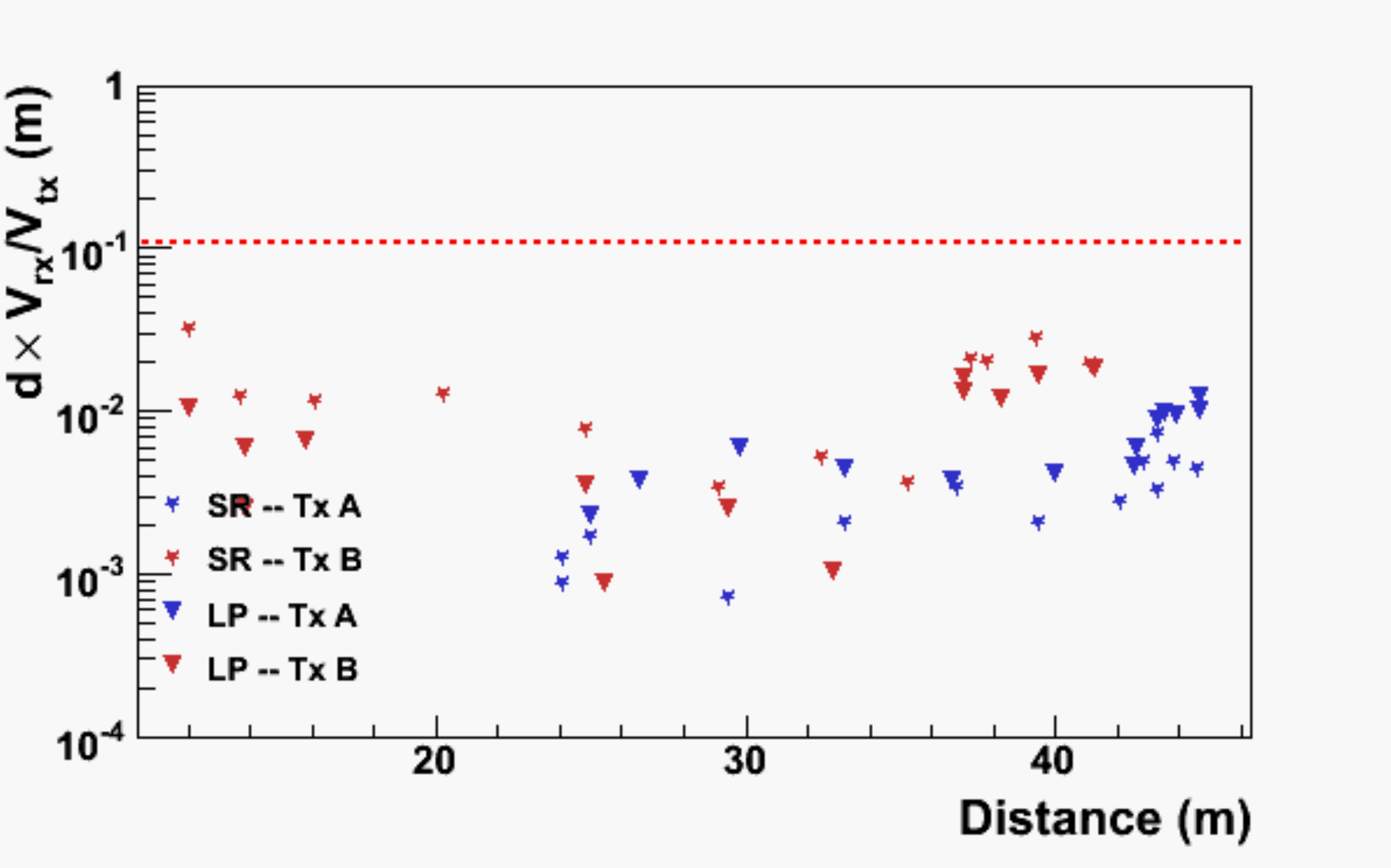}
    \caption{From the Hockley 200 measurements, peak voltages adjusted for 1/r loss as a function of distance.
\label{plot:2} }
\end{minipage}
\end{figure*}

First, the team repeated the 2002 Hockley measurement at the same
location with antennas
placed against the walls of salt, 
with the receiver placed at many positions around a corner 
(see Figure~\ref{plot:1}).  The integral of the power
spectrum in each antenna's frequency range was compared at each
distance to determine the attenuation lengths.
Recalling that the 2002 Hockley 
team observed no attenuation across the furthest
distance, the 2009 team
also observes an uptick in the signal strength 
near 40~m (see Figure~\ref{plot:2}).  This may be due to interference
between direct signals and ones reflected from the nearly parallel wall, leading to in essence
a waveguide effect.  This theory is supported by the poor quality
of the signals measured at the receivers.

The 2009 Hockley team also transmitted and received signals from within
boreholes, but only at 10~ft. depths.  These measurements show
field attenuation lengths in the 30-50~m range.  These are 
consistent with other measurements at Cote Blanche made at the mining levels,
where we believe fracturing in the salt leads to lossiness
not seen in the deep, undisturbed salt.

\section{Conclusions}
In~\cite{hockley}, the authors predict that a $10\times10\times10$ array
of dipole antennas with 50\% bandwidth could detect $\sim10$ BZ neutrinos
per year based on one typical model for cosmogenic neutrinos~\cite{ess}.  
This is based on an assumption of $L=300$~m 
at 300~MHz with $L\propto f^{-1}$.  If we
instead use the Cote Blanche borehole results, we predict of order
1 GZK event per year. 

It is becoming difficult to justify further {\it in situ} measurements
of radio/microwave attenuation lengths in salt due to their
difficulty and the strong possibility that the measured Cote Blanche
attenuation lengths in deep salt may be the best that can be found in
naturally occurring salt.

The best path forward for SalSA may be to utilize the $>100$~m  
attenuation lengths at low frequencies 
that we have measured at Cote Blanche, if 
the science-friendly mine management would allow us.
For example, SalSA 
could complement experiments in ice by working at lower frequencies,
where salt domes are shielded from galactic noise by rock above the salt.
Further studies need to be performed to explore this possibility.

\section{Acknowledgements}
We are grateful to Ben Straka and the other miners from the 
United Salt Corporation in Hockley, Texas for all the their
help with our successful 2009 trip to the Hockley mine,
making all of the
necessary arrangements including drilling boreholes for us into the
salt, and assisting the visiting team and ensuring their safety.
We would also like to thank the High Energy Physics Division of the 
U.S. Department of Energy,
the U.K. Particle Physics and Astronomy Research Council and
the Royal Society for funding this project.



\end{document}